\documentclass[journal]{IEEEtran}
\IEEEoverridecommandlockouts
\usepackage{cite}
\usepackage{amsmath,amssymb,amsfonts}
\usepackage{algorithmic}
\usepackage{algorithm}
\usepackage{graphicx}
\usepackage{textcomp}
\usepackage{xcolor}
\usepackage{subcaption}
\def\BibTeX{{\rm B\kern-.05em{\sc i\kern-.025em b}\kern-.08em
    T\kern-.1667em\lower.7ex\hbox{E}\kern-.125emX}}

\makeatletter
\long\def\@makecaption#1#2{\ifx\@captype\@IEEEtablestring%
\footnotesize\begin{center}{\normalfont\footnotesize #1}\\
{\normalfont\footnotesize\scshape #2}\end{center}%
\@IEEEtablecaptionsepspace
\else
\@IEEEfigurecaptionsepspace
\setbox\@tempboxa\hbox{\normalfont\footnotesize {#1.}~~ #2}%
\ifdim \wd\@tempboxa >\hsize%
\setbox\@tempboxa\hbox{\normalfont\footnotesize {#1.}~~ }%
\parbox[t]{\hsize}{\normalfont\footnotesize \noindent\unhbox\@tempboxa#2}%
\else
\hbox to\hsize{\normalfont\footnotesize\hfil\box\@tempboxa\hfil}\fi\fi}
\makeatother

\begin{document}
%
\title{Energy-Efficient UAV Communications: A Generalised Propulsion Energy Consumption Model}
%
%
%

\author{Xinhong~Dai,
        Bin Duo,~\IEEEmembership{Member,~IEEE},
        Xiaojun~Yuan,~\IEEEmembership{Senior~Member,~IEEE}, 
        and~Wanbin Tang,~\IEEEmembership{Member,~IEEE}
    \thanks{X. Dai, X. Yuan, and W. Tang are with the National Key Laboratory of Science and Technology on Communications, the University of Electronic Science and Technology of China, Chengdu 611731, China (e-mail: xhdai@std.uestc.edu.cn; xjyuan@uestc.edu.cn; wbtang@uestc.edu.cn). The corresponding author is Xiaojun Yuan.}
    \thanks{Bin Duo is with the National Laboratory of Science and Technology on Communications, University of Electronic Science and Technology of China, Chengdu 611731, China, and also with the College of Information Science and Technology, Chengdu University of Technology, Chengdu 610059, China (e-mail: duobin@cdut.edu.cn).}}
\maketitle
\begin{abstract}
This paper proposes a generalised propulsion energy consumption model (PECM) for rotary-wing ummanned aerial vehicles (UAVs) under the consideration of the practical thrust-to-weight ratio (TWR) with respect to the velocity, acceleration and direction change of the UAVs. To verify the effectiveness of the proposed PECM, we consider a UAV-enabled communication system, where a rotary-wing UAV serves multiple ground users as an aerial base station. We aim to maximize the energy efficiency (EE) of the UAV by jointly optimizing the user scheduling and UAV trajectory variables. However, the formulated problem is a non-convex fractional integer programming problem, which is challenging to obtain its optimal solution. To tackle this, we propose an efficient iterative algorithm by decomposing the original problem into two sub-problems to obtain a suboptimal solution based on the successive convex approximation technique. Simulation results show that the optimized UAV trajectory by applying the proposed PECM are smoother and the corresponding EE has significant improvement as compared to other benchmark schemes.
\end{abstract}


%
\IEEEpeerreviewmaketitle

\section{Introduction}
%
%
%
%
 Ummanned aerial vehicles (UAVs) are widely used in wireless communications due to their high flexibility, swift deployment capabilities and low cost characteristics \cite{yanmaz2018drone}. These advantages ensure that the UAVs can rapidly establish the high quality line-of-sight (LoS) communication channels with ground nodes \cite{zeng2016wireless}. Thus, the UAV-enabled communications will be widely used in typical 6G mobile communication scenarios in the future, such as natural disaster monitoring, border surveillance, emergency assistance, as well as relay and secure communications \cite{hayat2016survey,erdelj2017help,li2020reconfigurable,li2021robust}. Since most of the rotary-wing UAVs cannot be loaded with high-storage batteries, owing to their size and weight constraints \cite{gupta2015survey}, the lifetime of the UAVs is limited by their on-board payload. As such, the propulsion energy consumption issue cannot be neglected for UAV-enabled communication systems \cite{yan2020uav}. 
\par Since the UAVs' trajectories play a critical role in their propulsion energy consumption, there have been quite a few works on the modeling of their propulsion energy consumption and trajectory designs. The authors in \cite{liu2017power} derived a multi-rotor UAV power consumption model based on the helicopter theory, by assuming that the UAV keeps a steady flying state. In \cite{zeng2019energy}, the authors aimed to minimize the total energy consumption and proposed a propulsion energy consumption model (PECM) by only considering velocity of the rotary-wing UAV. The PECM in \cite{zeng2019energy} was improved in \cite{yang2019energy} by introducing the UAV's accerleration under the consideration of straight forward level flight. In \cite{gao2021energy}, the authors generalized the PECM by considering the centrifugal acceleration and kinetic energy change over the given time duration, and extended the flight scenario to an arbitrary two-dimensional (2D) level flight. Based on the above proposed PECMs, there are also many works focusing on the trajectories designs of the UAVs to improve the energy efficiency (EE) of the UAV-enabled communication systems \cite{duo2020energy,zhan2019energy,gong2018flight}. 
\par The optimized UAV trajectories of above works based on conventional PECMs show that the UAVs make frequent and sharp turns, especially when flying above the ground users. In practice, these flight behaviours undoubtedly cause additional propulsion energy consumption of the UAVs. As a key fuselage parameter in the PECM of rotary-wing UAVs, the thrust-to-weight ratio (TWR) is the ratio of thrust generated by fuselage propeller to their own weight, which is highly related to the UAV flight status such as velocity, acceleration and direction change \cite{filippone2006flight}. However, the relation between the TWR and the flight status is usually oversymplified in the existing works (e.g., by letting the TWR equal to 1 in \cite{zeng2019energy}) and cannot characterize UAV's real propulsion energy consumption. As such, it is of pressing importance to establish a generalised PECM with more accurate characterization of the relation between the TWR and the flight status, based on which more energy-efficient UAV trajectory design can be realized.
\par Motivated by the above, this paper proposes a generalised PECM of the rotary-wing UAV, by analyzing the TWR with respect to (w.r.t.) the velocity, acceleration and direction change of the UAV. We jointly optimize the user scheduling and trajcectory design to maximize the EE of the typical UAV-enabled communication system. Specifically, we formulate a fractional integer programming problem based on the proposed PECM, which is challenging to solve. To tackle the non-convexity of the original problem, we propose an efficient iterative algorithm and decompose the problem into two subproblems to obtain a suboptimal solution based on the alternative optimization (AO) and Dinkelbach method with successive convex approximation (SCA) technique. The simulation results show that the optimized trajectory based on the proposed PECM is smoother, and the UAV achieves higher EE at different flying durations as compared to the benchmark schemes. This indicates that the optimized trajectory of the UAV with the proposed PECM is more energy-efficient, which is based on the comprehensive analysis of the TWR.

\section{Generalised PECM}
In this section, we first take the force analysis of rotary-wing UAVs to obtain an expression of the TWR w.r.t. the velocity, acceleration and direction change. Then, we derive the proposed PECM.

\begin{figure}[h]
    \centering
    \includegraphics[width=6.5cm]{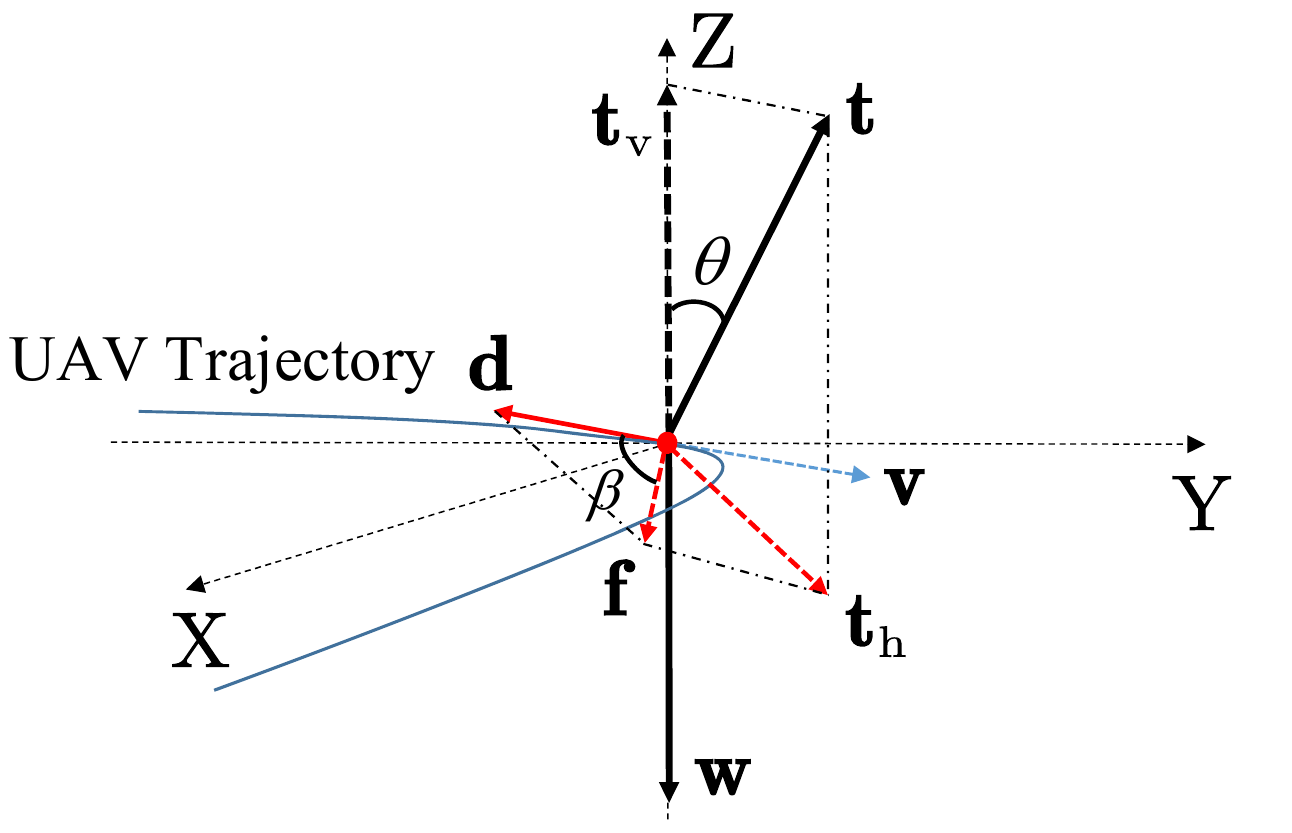}
    \caption{ Horizontal force analysis on a rotary-wing UAV}
    \label{Force all}
\end{figure} 

\par To obtain an expression of the TWR, the horizontal force of the rotary-wing UAV is analyzed as shown in Fig. \ref{Force all}, which includes the following forces: (i) $\mathbf{t}$: rotor thrust, which is perpendicular to the disc plane and directed upward, and $\mathbf{t}_{\rm h}$ and $\mathbf{t}_{\rm v}$ represents the component of UAV's thrust in the horizontal direction and the vertical direction, respectively; (ii) $\mathbf{d}$: fuselage drag, which is in the opposite direction of the aircraft velocity \cite{bramwell2001bramwell}, and
\begin{align}
\mathbf{d} = -\frac{1}{2}\mathit{\rho S_{FP}}{\|\mathbf{v}}\|^2\mathbf{e}_{\rm v},
\label{D}
\end{align}
where $\mathbf{v}$ denotes the instantaneous velocity of the UAV, and its direction is along the tangent direction of the UAV trajectory, $\mathit{S_{FP}}$ denotes the fuselage equivalent flat plate area, and $\mathbf{e}_{\rm v}$ is the unit direction vector of $\mathbf{v}$; (iii) $\mathbf{w}$: aircraft weight, and $\|\mathbf{w}\|=mg$, where $m$ is the total mass of the UAV and $g$ is the vertical downward gravitational acceleration.
\par From Fig. \ref{Force all}, we first denote $\theta$ as the tilt angle of the rotor disc, and define $\|\mathbf{t}_{\rm h}\|\triangleq\|\mathbf{t}\||\sin{\theta}|$ and $\|\mathbf{t}_{\rm v}\|\triangleq\|\mathbf{t}\||\cos{\theta}|$. Then, based on the law of cosines, we have
\begin{align}
\|\mathbf{t}\||\sin\theta| = \sqrt{\|\mathbf{f}\|^2+\|\mathbf{d}\|^2-2\|\mathbf{f}\|\|\mathbf{d}\|\cos\beta,}
\label{T_sin_theta_origin}
\end{align}
where $\mathbf{f}=m\mathbf{a}$ is the equivalent synthetic force caused by $\mathbf{t}_{\rm h}$ and $\mathbf{d}$, $\mathbf{a}$ denotes the instantaneous acceleration of the UAV, $\beta$ represents the angle between the direction of $\mathbf{f}$ and $\mathbf{d}$, and $\cos\beta = -\cos(\pi-\beta) = - \frac{\mathbf{a}\cdot\mathbf{v}}{\|\mathbf{a}\|\|\mathbf{v}\|}$. From the force balance in the vertical direction, we can obtain
\begin{align}
\label{T_cos_theta_origin}
\mathit{\|\mathbf{t}\||\cos\theta|=\|\mathbf{w}\|}.
\end{align}

\par Denote by $c=\mathbf{a}\cdot\mathbf{v}$ the direction change of the UAV flight. From \eqref{D}-\eqref{T_cos_theta_origin}, the TWR $\kappa$ can be written as
\begin{align}
\label{new_kappa}
\kappa(\mathbf{a},\mathbf{v}) = \sqrt{1+\frac{4m\|\mathbf{a}\|^2+\rho^2 S_{FP}^2\|\mathbf{v}\|^4+4m\rho S_{FP}c\|\mathbf{v}\|}{4\|\mathbf{w}\|^2}}.
\end{align}
We then obtain a generalised PECM w.r.t. the velocity, accerleration and direction change of the rotary-wing UAVs as
\begin{align}
\label{P_kappa}
\mathit{P}(\mathbf{a},\mathbf{v})
= & \mathit{P_{i}}\kappa(\mathbf{a},\mathbf{v})\sqrt{\left(\kappa^2(\mathbf{a},\mathbf{v})+\frac{\|\mathbf{v}\|^{4} }{4v_{\rm 0}^{4} }\right)^{\frac{1}{2} } -\frac{\|\mathbf{v}\|^{2}}{2v_{\rm 0}^{2}}}\nonumber\\
&+\mathit{P_{\rm0}}\phi(\mathbf{v})+\psi(\mathbf{v}),    
\end{align}
where $\phi(\mathbf{v}) = (1+\frac{3\|\mathbf{v}\|^2}{U_{\rm tip}^2})$ and $\psi(\mathbf{v}) = \frac{1}{2}d_0\rho sA\|\mathbf{v}\|^3$, $\mathit{P_{\rm 0}}$ and $\mathit{P_{i}}$ represent the \textit{blade profile power} and the \textit{induced power} in hovering status, respectively, and $\psi(\mathbf{v})$ represents the \textit{parasite power} of UAV. $\mathit{U_{\rm tip}}$ denotes the tip velocity of the rotor blade, $\mathit{v_{\rm0}}$ is the mean rotor induced velocity in hover, $\mathit{d_{\rm0}}$ and $\mathit{s}$ are the fuselage drag ratio and rotor solidity, respectively, and $\mathit{\rho}$ and \emph{A} denote the air density and the rotor disc area, respectively. Other parameters are detailed in  [\citenum{zeng2019energy}, Table \uppercase\expandafter{\romannumeral1}].
\par Note that different from the approximation of $\kappa$ in \cite{zeng2019energy}, the PECM proposed here characterizes the parctical flight status of the rotary-wing UAV more accurately, which helps to explore more energy-efficient UAV trajectory designs.

\section{System Model and Problem Formulation}
In this section, we consider a UAV-enabled broadcast communication system, and aim to improve the EE of the UAV based on the proposed PECM in \eqref{P_kappa}. Specifically, a rotary-wing UAV flies at a fixed altitude denoted as \textit{H}, and communicates with $K > 1$ ground users denoted by the set $\mathcal{K} = \{1,\dots,K\}$. We assume that each ground user \emph{k} is located at $\mathbf{g}_k = [x_k,y_k]\in \mathbb{R}^{2\times1}, k \in \mathcal{K}$. For tractablity, we denote \emph{T} as the total time in which the UAV serves all users and divide it into \emph{N} time slots with equal length, indexed by $n=1,\dots,N$, and the elemental slot length $\delta_t = \frac{T}{N}$ is set sufficiently small. The UAV's horizontal trajectories over \emph{T} can be approximated by the position sequences $\mathbf{Q}=\{\mathbf{q}[n] = [x[n],y[n]]^T, \forall n\}$. We denote $\Omega = V_{max}\delta_t$ as the UAV's maximum horezontal movement with each time slot, where $V_{max}$ is the UAV's maximum velocity in meter/second (m/s). We assume that the UAV needs to return to its initial location by the end of \textit{T}, and the UAV trajectory satisfies the following constraints
\begin{align}
\label{constaint_v}
\|\mathbf{q}[n+1]-\mathbf{q}[n]\|^2 \le \Omega^2,\quad n = 0,\dots,N-1,
\end{align}
\begin{align}
\label{constraint_start_end }
\mathbf{q}[0]=\mathbf{q}[N].
\end{align}
It is assumed that the comunication links between the UAV and the users are dominated by LoS links [\citenum{lin2018sky}]. Thus, the channel power gain follows the free-space path loss model expressed as $h_k[n]=\rho_0 d^{-2}_k[n] = \frac{\rho_0}{H^2+\|\mathbf{q}[n]-\mathbf{g}_k\|^2}$, where $\rho_0$ denotes the channel power gain at the unit reference distance, and $d_k[n] = \sqrt{H^2+\|\mathbf{q}[n]-\mathbf{g}_k\|^2}$ is the distance from the UAV to user \textit{k} in time \textit{n}. 
\par We assume that the UAV sends the signal with a constant power $\mathit{P}_0$, which can be ignored in the total system energy consumption because the communication-related energy is much smaller than the propulsion energy of UAVs \cite{duo2020energy}. We assume that all users share the same frequency band \textit{B} over the duration $T$ in second ($s$), and communicate via the time-division multiple access (TDMA). We define the user scheduling variable as $\mathbf{A}=\{\alpha_k[n],\forall k,n\}$, in which $\alpha_k[n]$ indicates that user \textit{k} is served by the UAV in time slot \textit{n} if $\alpha_k[n] = 1$; otherwise, $\alpha_k[n] = 0$. As such, the achievable rate of user \emph{k} in time slot \emph{n} can be expressed as
\begin{align}
R_k[n] = \alpha_k[n]\log_2(1+\gamma_k[n]),
\end{align}
where $\gamma_k[n] = \frac{P_0 h_k[n]}{\sigma^2} = \frac{\gamma_0}{H^2+\|\mathbf{q}[n]-\mathbf{g}_k\|^2}$ is the signal-to-noise ratio (SNR) at user \emph{k} in time slot \textit{n}, $\gamma_0 = \frac{P_0 \rho_0}{\sigma^2}$, and $\sigma^2$ is the power of the additive white Gaussian noise (AWGN) at the receiver. Based on the discretization for UAV trajectory, the velocity and acceleration of the UAV at time slot \emph{n} can be expressed as $\mathbf{v}[n]=\frac{\mathbf{q}[n+1]-\mathbf{q}[n]}{\delta_t}$ and $\mathbf{a}[n]=\frac{\mathbf{q}[n+2]+\mathbf{q}[n]-2\mathbf{q}[n+1]}{\delta_t^2}$, respectively. Therefore, the PECM in time slot \textit{n} can be described as 
\begin{align}
\mathit{P[n]} 
\label{P_kappa_dis}
& = \mathit{P_{\rm 0}}\phi[n] +\psi[n]\nonumber\\
&+ \mathit{P_{i}}\kappa[n]\sqrt{\left((\kappa[n])^2+\frac{\|\mathbf{v}[n]\|^4}{4v_{\rm0}^{4} } \right)^{\frac{1}{2} } -\frac{\|\mathbf{v}[n]\|^2}{2v_{\rm0}^{2}}},    
\end{align}
where $\phi[n] = (1+\frac{3\|\mathbf{v}[n]\|^2}{U^2_{\rm {tip}}})$ and $\psi[n] = \frac{1}{2}d_0\rho sA\|\mathbf{v}[n]\|^3$, $\kappa[n]$ is the exprssion of the TWR in time slot \textit{n} based on \eqref{new_kappa}. To maximize the EE of the UAV-enabled communication system over \textit{N} time slots, we jointly optimize the UAV trajectory variable $\mathbf{Q}$ and the user scheduling variable $\mathbf{A}$. The optimization problem can be formulated as
 \begin{subequations}\label{problem_origin}
 \begin{align}
 \underset{\mathbf{Q},\mathbf{A},R_{min}}{\max}\quad &\frac{R_{min}}{\sum_{n=1}^{N}P[n]}\label{function_origin}\\
s.t.\quad&\label{constraint_alpha_0,1}
\alpha_k[n]\in{\{0,1\}},\quad \forall k,n, \\
& \label{constraint_alpha<=1}
\sum_{k=1}^{K}\alpha_k[n]\le1,\quad \forall n,\\
& R_{min}\le\sum_{n=1}^{N}R_k[n],\quad \forall k,\label{constraint_fairness}\\
&\eqref{constaint_v},\eqref{constraint_start_end },
\end{align}
\end{subequations}
where $R_{min}=B\underset{k\in \mathcal{K}}{\min} \sum_{n=1}^{N}R_k[n]$ is defined as the minimum average rate among all users over $N$ time slots, and constraint \eqref{constraint_alpha<=1} indicates that the UAV serves at most one user in each time slot. Since $P[n]$ is non-convex w.r.t. $\mathbf{q}[n]$, and constraint \eqref{constraint_alpha_0,1} is an integer constraint, problem \eqref{problem_origin} is a mixed-integer non-convex fractional program, which is chanllenging to solve in general.

\section{Proposed Algorithm}
\par In this section, we propose an efficient iterative algorithm to solve problem \eqref{problem_origin} approximately by applying the AO and SCA methods. Specifically, problem \eqref{problem_origin} is decomposed into two sub-problems, where the user scheduling variable $\mathbf{A}$ and the UAV trajectory variable $\mathbf{Q}$ are optimized alternately. In this way, we obtain a suboptimal solution to problem \eqref{problem_origin} when the algorithm converges to a given threshold $\epsilon >$ 0.
\subsection{User Scheduling Optimization}
\par For any given $\mathbf{Q}$, the sub-problem is equivalent to maximize $R_{min}$ in \eqref{function_origin}. To make the subproblem more tractable, we relax the binary variables in \eqref{constraint_alpha_0,1} into continous variables, which yields the following subproblem:
 \begin{subequations}\label{subproblem_alpha_relax}
 \begin{align}
 \underset{\mathbf{A},R_{min}}{\max}\quad &R_{min}\label{function_alpha_relax}\\
s.t.\quad&0\le \alpha_k[n]\le 1,\quad \forall k,n,\label{constraint_alpha_relax_0,1}\\
&\sum_{k=1}^{K}\alpha_k[n]\le1,\quad \forall n,\\
&R_{min}\le\sum_{n=1}^{N}\alpha_k[n]\log_2(1+\gamma_k[n]),\quad \forall k.
\end{align}
\end{subequations}
Since subproblem \eqref{subproblem_alpha_relax} is a linear program, it can be sloved efficiently by CVX \cite{cvx}.
\subsection{UAV Trajectory Optimization}
For any given $\mathbf{A}$, the UAV trajectory variable $\mathbf{Q}$ can be optimized by solving the following problem:
 \begin{subequations}\label{subproblem_origin_q}
 \begin{align}
 \underset{\mathbf{Q},R_{min}}{\max}\quad &\frac{R_{min}}{\sum_{n=1}^{N}P[n]}\label{function_origin_q}\\
s.t. \quad & R_{min}\le\sum_{n=1}^{N}R_k[n],\quad \forall     k,\label{constraint_fairness}\\
& \eqref{constaint_v},\eqref{constraint_start_end }.
\end{align}
\end{subequations}
\par Problem \eqref{subproblem_origin_q} is non-convex. Firstly, to tackle the non-convexity of constraint \eqref{constraint_fairness}, we introduce slack variable $\mathbf{G}=\{G_k[n],\forall k,n\}$, in which $G_k[n]$ satisfies
\begin{equation}
  G_k[n]\ge H^2+\|\mathbf{q}[n]-\mathbf{g}_k\|^2, \quad \forall n,k. \label{constraint_Gk}
\end{equation}
Thus, $R_k[n]$ can be written as $R_k[n] = \alpha_k[n]\log_2(1+\frac{\gamma_0}{G_k[n]})$. By using the first-order Taylor expansion, $R_k[n]$ can be replaced by its convex lower bound, at given local points denoted by $\mathbf{G}^t=\{G_k^t[n],\forall k,n\}$ in the \emph{t}-th iteration, i.e.,
\begin{align}
\alpha_k[n]\log_2(1+\frac{\gamma_0}{G_k[n]}) \ge R_k^{lb}[n], 
\end{align}
where $R_k^{lb}[n] = \alpha_k[n](\log_2(1+\frac{\gamma_0}{G_k^t[n]})-\frac{\gamma_0(G_k[n]-G_k^t[n])}{\ln{2}(G_k[n]+\gamma_0)G_k^t[n]})$. Note that \eqref{constraint_Gk} must hold with equalities to obtain the optimal solution to problem \eqref{subproblem_origin_q}. Otherwise $G_k[n]$ can be increased to further reduce the objective value.
\par Next, we deal with the non-convexity of $P[n]$ in the objective function \eqref{function_origin_q}, and the specific steps are as follows: (i) Since $c\|\mathbf{v}\|\in[-\|\mathbf{a}\|\|\mathbf{v}\|^2,\|\mathbf{a}\|\|\mathbf{v}\|^2]$, and it is easy to verify that $P[n]$ increases with the increasing of $\kappa[n]$, we can approximate $c\|\mathbf{v}\|$ as $\|\mathbf{a}\|\|\mathbf{v}\|^2$ to obtain the upper bounds of $\kappa[n]$ and $P[n]$, denoted as $\hat{\kappa}[n]= \sqrt{1+\frac{(2m\|\mathbf{a}[n]\|+\rho S_{FP}\|\mathbf{v}[n]\|^2)^2}{4(mg)^2}}$ and $\hat{P}[n]$, respectively; (ii) we introduce the slack variable $\mathbf{S} = \{S[n]\}_{n=1}^N$ such that
\begin{equation}
\label{S[n]}
S[n]\ge \hat{\kappa}[n]\sqrt{\left((\hat{\kappa}[n])^2+\frac{\|\mathbf{v}[n]\|^4 }{4v_{0}^{4} } \right)^{\frac{1}{2}}-\frac{\|\mathbf{v}[n]\|^2}{2v_{0}^{2}}},
\end{equation}
which is still non-convex for $\forall n$; (iii) to tackle the non-convexity of \eqref{S[n]}, we further introduce the slack variables $\mathbf{L} = \{L[n]\}_{n=1}^N$ and $\mathbf{M} = \{M[n]\}_{n=1}^N$, which satisfy
$L[n]\ge \hat{\kappa}[n]$ and $M[n]\ge ((\hat{\kappa}[n])^2+\frac{\|\mathbf{v}[n]\|^4 }{4v_{0}^{4} } )^{\frac{1}{2} } -\frac{\|\mathbf{v}[n]\|^2}{2v_{0}^{2}}$. Based on the steps above, we have the following constraints in any time slot \emph{n}:
\begin{equation}
\label{constraint_t}
1+\left(\frac{\|\mathbf{a}[n]\|}{g}+\frac{\rho S_{FP}\|\mathbf{v}[n]\|^2}{2mg}\right)^2 \le L^2[n],
\end{equation}
\begin{equation}
\label{constraint_M}
\frac{L^2[n]}{M[n]} \le M[n] + \frac{\|\mathbf{v}[n]\|^2}{v^2_0},
\end{equation}
\begin{equation}
\label{constraint_S}
L^2[n]\le \frac{S^2[n]}{M[n]}.
\end{equation}
Note that \eqref{constraint_S} holds wtih equality for $\forall n$ to obtain the optimal solution. Otherwise \emph{t}[n] and \emph{M}[n] can be increased and so can \emph{S}[n] to further reduce the objective value of problem \eqref{problem_origin}. 
\par For \eqref{constraint_t}, we set $\mu[n]=\frac{\|\mathbf{a}[n]\|}{g}+\frac{\rho S_{FP}\|\mathbf{v}[n]\|^2}{2mg}$ which is convex w.r.t. $\mathbf{q}[n]$. To convert the inequalities \eqref{constraint_t}-\eqref{constraint_S} into convex constraints, we apply the first-order Taylor expansion to obtain their lower bounds, at given local points denoted by $\mathbf{Q}^t = \{\mathbf{q}^t[n]\}_{n=1}^N$, $\mathbf{S}^t = \{S^t[n]\}_{n=1}^N$, $\mathbf{L}^t = \{L^t[n]\}_{n=1}^N$, and $\mathbf{M}^t = \{M^t[n]\}_{n=1}^N$ in the \emph{t}-th iteration, respectively. Next, we define $\mathbf{v}^t[n]=\frac{\mathbf{q}^t[n+1]-\mathbf{q}^t[n]}{\delta_t}$, and have the following convex constraints in each time slot \emph{n}:
\begin{align}
\label{constraint_t_taylor}
1+\mu^2[n]\le (L^t[n])^2+2L^t[n](L[n]-L^t[n]),
\end{align}
\begin{align}
\label{constraint_M_taylor}
\frac{L^2[n]}{M[n]}\le M[n]-\frac{\|\mathbf{v}^t[n]\|^2}{v_0^2}+2\frac{(\mathbf{v}^t[n])^T\mathbf{v}[n]}{v_0^2},
\end{align}
\begin{align}
\label{constraint_S_taylor}
L^2[n] \le &\frac{(S^t[n])^2}{M^t[n]}+\frac{2S^t[n](S[n]-S^t[n])}{M^t[n]}\nonumber\\
&-\left(\frac{S^t[n]}{M^t[n]}\right)^2(M[n]-M^t[n]).
\end{align}
\par Based on the above derived lower bounds, problem \eqref{subproblem_origin_q} can be reformulated as
 \begin{subequations}\label{subproblem_q}
 \begin{align}
 \underset{\mathbf{Q},\mathbf{G},\mathbf{S},\mathbf{M},\mathbf{L},R_{min}}{\max}\quad &\frac{R_{min}}{\sum_{n=1}^{N}
 (P_{0}\phi[n] + P_{i}S[n]+\psi[n])}\label{function_q}\\
s.t.\quad&R_{min}\le\sum_{n=1}^{N}R_k^{lb}[n],\quad \forall k,\\
&\eqref{constaint_v},\eqref{constraint_start_end },\eqref{constraint_Gk},\eqref{constraint_t_taylor}-\eqref{constraint_S_taylor}.
\end{align}
\end{subequations}
\par Problem \eqref{subproblem_q} is a quasi-convex optimization problem since \eqref{function_q} includes linear numerator and convex denominator, and all constraints of \eqref{subproblem_q} are convex. As such, it can be solved optimally and efficiently via fractional programming techniques, e.g., the Dinkelbach’s algorithm [\citenum{shen2018fractional}].
\begin{figure*}[htb]
\centering
\begin{minipage}{0.32\linewidth}
\centering
\includegraphics[width=6.5cm]{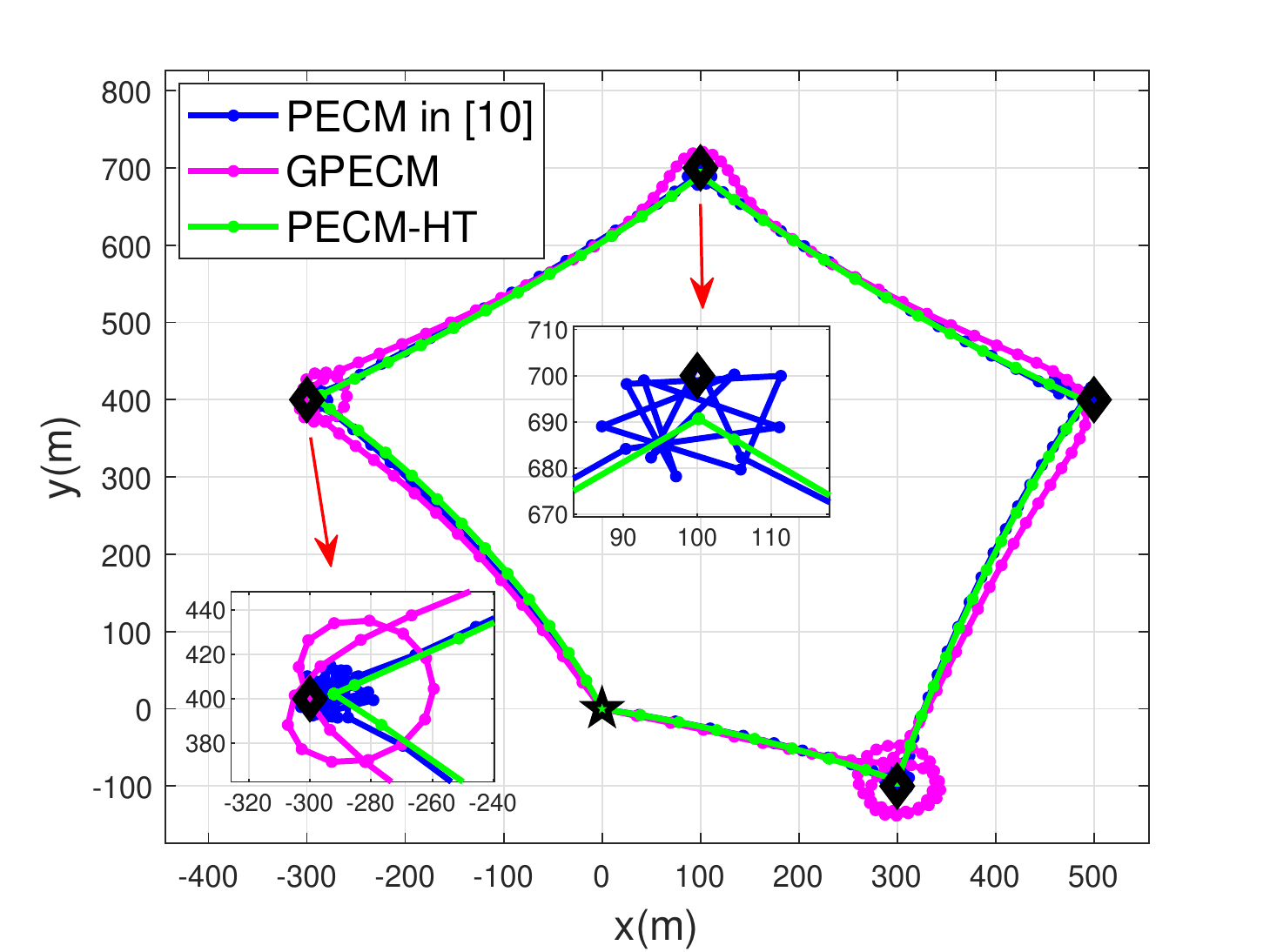}
\caption{Trajectories}
\label{traj}
\end{minipage}
\begin{minipage}{0.32\linewidth}
\centering
\includegraphics[width=6.5cm]{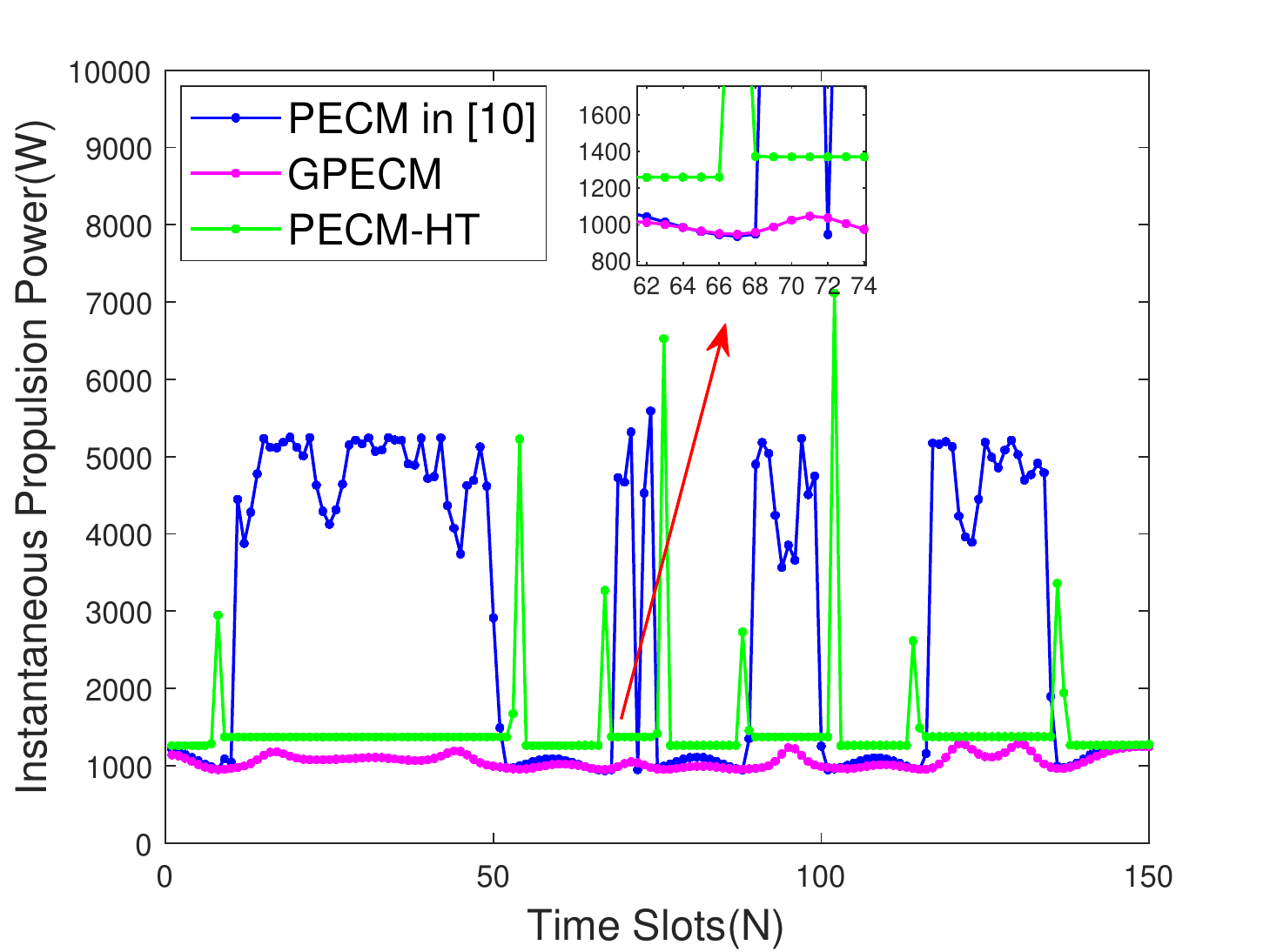}
\caption{Instantaneous propulsion power}
\label{power}
\end{minipage}
\begin{minipage}{0.32\linewidth}
\centering
\includegraphics[width=6.5cm]{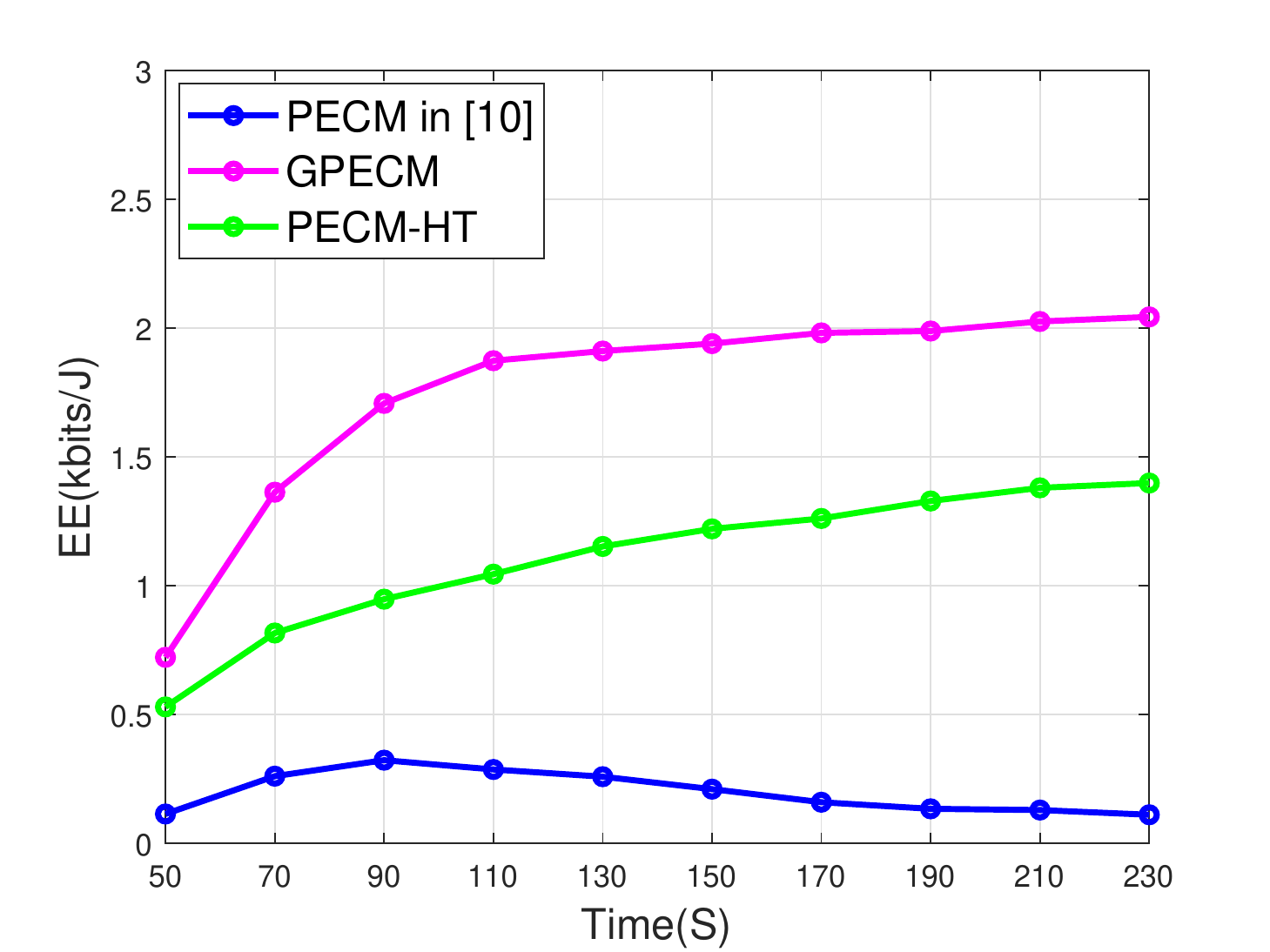}
\caption{EE comparision}
\label{EE}
\end{minipage}
\end{figure*}
 
 \subsection{Overall Algorithm and Computational Complexity}
 \par Based on the previous two subsections, we summarize the proposed algorithm as follows: (i) We denote iteration index as \textit{r} to optimize two subproblems alternately, and initialize $\mathbf{Q}^0,\mathbf{A}^0$; (ii) we alternately optimize subproblem \eqref{subproblem_alpha_relax} with given $\mathbf{Q}^r$ and \eqref{subproblem_q} with given $\mathbf{A}^{r+1}$ in the \textit{r}-th iteration; (iii) we take the binary decision for optimized $\mathbf{A}$ to make it integerized, when the difference between the objective values of two subproblems converges within a pre-specified precision $\epsilon>0$. By using the proposed algorithm, problem \eqref{problem_origin} can be approximately tackled by iteratively updating the optimization variables based on the optimal solutions to two subproblems \eqref{subproblem_alpha_relax} and \eqref{subproblem_q}. The relaxation of $\mathbf{A}$ and constraints \eqref{constraint_t_taylor}-\eqref{constraint_S_taylor} guarantee the convergence as analyzed in \cite{zhang2019securing}. Moreover, the asymptotic computational time complexity of the proposed algorithm is
$\mathcal{O}(N^{3.5}\log(1/\epsilon))$, which means that the suboptimal solution can be computed in polynomial time, and thus can be efficiently implemented in practical UAV-enabled communication systems.
 
 \section{Numercial Results}
 \par In this section, we show the simulation results to compare the EE maximization by applying the proposed generalised PECM (denoted as GPECM) with two benchmark schemes: 1) EE maximization by applying the PECM in \cite{zeng2019energy}; 2) average rate maximization among users based on the heuristic trajectories (denoted as PECM-HT). In this scheme, we assume that the UAV flies at its maximum velocity among four users that are randomly distributed within a 2D area, and keeps hovering after reaching above each user. The UAV's initial and final positions are set as $\mathbf{q}[0]=\mathbf{q}[N] = [0,0]^T$ m, and four ground users' horizontal locations are $[300,-100]^T$ m, $[500,400]^T$ m, $[100,700]^T$ m and $[-300,400]^T$ m, respectively. The other parameters are set as $H$ = 100 m, $W = 50$ N, $P_0$ = 20 dBm, $V_{max}$ = 40 m/s, $\delta_t$ = 1 s, $\rho_0 = -60$ dBm, $B=1$ MHz, $\sigma^2=-90$ dBm, and $\epsilon = 10^{-4}$. The settings of the remaining parameters in GPECM refer to [\citenum{zeng2019energy}, Table \uppercase\expandafter{\romannumeral1}].
 \par Fig. \ref{traj} shows the optimized trajectories of the UAV based on different schemes with given $T$ = 150 s. We can find that all trajectories are generally similar when the UAV flies between each user, but present significantly different trajectories when flying above each user. Specifically, the UAV with the GPECM flies above each user in a smoother trajectory. By contrast, the UAV with the PECM in \cite{zeng2019energy} flies above each user in a manner of frequent and sharp turns in a small range for higher EE. Under the consideration of the UAV's actual flight status, flying in this way would undoubtedly bring about the additional propulsion energy consumption for UAVs, as verified in Fig. \ref{power}. 
 \par Fig. \ref{power} shows the changes in instantaneous propulsion power of the UAV based on different schemes over \textit{N} time slots. Specifically, compared with the other two benchmark schemes, the instantaneous propulsion power of the proposed scheme is lowest during the entire flying process, since the optimized trajectory based on the GPECM is smoother, and thus is more energy conservation. However, the instantaneous propulsion power of the UAV with the PECM in \cite{zeng2019energy} is much higher than that of the other two schemes, especially when the UAV keeps turning above each user. The main reason is that the PECM in \cite{zeng2019energy} did not fully consider the generalised $\kappa$ w.r.t. the acceleration and direction change of the UAV. Therefore, the UAV with the PECM in \cite{zeng2019energy} flies in a manner of frequent and sharp turns above each user, which causes addtional propulsion energy consumption. The instantaneous propulsion power of the UAV with the PECM-HT is relatively stable, due to its constant volecity in the hovering and flying status. We also notice that although PECM-HT is a simpler scheme, it is not the most energy-efficient flight scheme for the UAV.
 \par Fig. \ref{EE} shows the obtained EE based on different schemes in different \textit{T}. As expected, the EE of the UAV with the GPECM significantly outperforms other benchmark schemes. Specifically, the EE based on the PECM in \cite{zeng2019energy} are significantly lower than other schemes, which indicates the importance of considering the generalised $\kappa$ as well as the generalised PECM for the UAVs. In addtion, the EE of the UAV with the GPECM is significantly higher than with the PECM-HT, which demonstrates the necessity for the trajectory designs based on the proposed PECM.

\section{Conclusion}
To characterize the PECM of rotary-wing UAVs in actual flight status more accurately, this paper proposed a generalised PECM by analyzing the TWR w.r.t. the velocity, accleration, and direction change of the UAVs. To verify the effectiveness of such model, we considered a typical UAV-enabled mutiple access communication. To maximize the EE of the UAV, we jointly designed the user scheduling and UAV trajectory variables. The formulated problem is a non-convex integer fractional program, which is challenging to solve in general. Thus, we proposed an efficient iterative algorithm to slove it. Numerical results showed that the UAV trajectory obtained by applying the proposed GPECM is smoother than the benchmark scheme with $\kappa = 1$, and the UAV saves more propulsion energy to achieve higher EE. 

%




\ifCLASSOPTIONcaptionsoff
  \newpage
\fi



%


\bibliographystyle{IEEEtran} 
\bibliography{reference}

%








\end{document}